# Quantum Convolutional Neural Network with Nonlinear Effects and Barren Plateau Mitigation

Pei-Kun Yang

E-mail: peikun@isu.edu.tw, peikun6416@gmail.com

## Abstract

Quantum neural networks (QNNs) leverage quantum entanglement and superposition to enable large-scale parallel linear computation, offering a potential solution to the scalability limits of classical deep learning. However, their practical deployment is hampered by two key challenges: the lack of intrinsic nonlinear operations and the barren plateau phenomenon. We propose a quantum convolutional neural network (QCNN) architecture that simultaneously addresses both issues. Nonlinear effects are introduced via orthonormal basis expansions of power series, while barren plateaus are mitigated by directly parameterizing unitary matrices rather than stacking multiple parameterized gates. Our design further incorporates quantum analogs of convolutional kernels and strides for scalable circuit construction. Experiments on MNIST and Fashion-MNIST datasets achieve 99.0% and 88.0% test accuracy, respectively. Consistency between PyTorch-based matrix simulation and Qiskit-based quantum circuit simulation validates the physical fidelity of the model. These results demonstrate a flexible and effective quantum architecture that faithfully integrates classical convolutional mechanisms into a quantum framework, paving the way for practical and expressive QNNs.

## Introduction

Classical artificial neural networks have become a cornerstone of modern computing. They're used in everything from image recognition to autonomous vehicles because they can learn patterns by adjusting parameters across layers [1-3]. These layers depend on matrix operations, and as networks become deeper or data becomes more complex, the computational load can grow rapidly. This challenge has led researchers to consider whether quantum computing could offer a fundamentally different and more efficient approach.

Quantum computers utilize features such as superposition and entanglement, which enable them to represent and process all possible quantum states simultaneously [4-7]. With $n$ qubits, a quantum system can describe $2^n$ distinct states. Each state has an address written as $s_{n-1}\ldots s_{p+1}s_p s_{p-1}\ldots s_1 s_0$, where each $s_m$ is either 0 or 1. Now, suppose we apply a Hadamard, Rx, or Ry gate to the $p^{th}$ qubit in the circuit. This gate creates a linear combination of two quantum states: one where the $p^{th}$ bit is 0, and one where it is 1. The corresponding addresses are $s_{n-1}\ldots s_{p+1}0s_{p-1}\ldots s_1 s_0$ and $s_{n-1}\ldots s_{p+1}1s_{p-1}\ldots s_1 s_0$. The important point here is that applying this single gate changes the amplitudes of all output quantum states. This is very different from classical systems, where an operation usually only affects a limited portion of memory. In a quantum system, even a single-qubit gate requires recalculating all amplitudes in the state vector to reflect the new global quantum state.

However, while this global sensitivity makes quantum systems highly effective for applying the same operation across all states in parallel, it also introduces significant challenges. When each quantum state needs to undergo a different transformation based on certain conditions, such as in conditional logic, the circuit design becomes more complex. Applying a gate only when a qubit is in a specific state requires multi-controlled operations, which are increasingly difficult to implement as the number of qubits grows.

In other cases, the difficulty lies not in conditional behavior, but in combining information from different quantum states. For example, adding two specific quantum states may seem simple at first. However, this is a different kind of challenge that requires constructing gates capable of merging amplitudes across selected states. Implementing such operations typically demands a sequence of complex controlled gates, and decomposing them into elementary gates further increases the depth and resource cost of the circuit.

By contrast, classical processors like CPUs and GPUs handle these types of operations with ease. Arithmetic tasks such as addition and subtraction can be carried out directly and efficiently, with precise control over individual memory locations. Classical hardware excels at performing localized, state-specific updates, whereas quantum circuits must carefully manage global state evolution through reversible and unitary operations. This fundamental difference means that while quantum computers offer great advantages for uniform, parallel processing, they are less suited for operations that require targeted manipulation of specific data paths.

Heterogeneous computing is becoming an essential paradigm in modern system architecture. In this approach, different types of processors are used based on the nature of each task, allowing workloads to be divided across CPUs, GPUs, and other specialized units to improve efficiency and performance. PyTorch offers one practical example of this trend, as it allows users to assign specific operations to either CPUs or GPUs depending on their computational needs. However, this kind of flexibility is not limited to a single framework. Looking ahead, computational tasks may be distributed more dynamically across various processors, including quantum processing units (QPU). These devices might not even be located within the same physical machine. Instead, tasks can be sent over a network, processed remotely, and the results returned to the main system. In this distributed computing model, each processor handles the part of the workload that best matches its capabilities. This cooperative structure brings together classical and quantum resources into a unified and adaptable system, to optimize overall performance and scalability.

In recent decades, researchers have studied molecular interactions inside cells using methods from statistical thermodynamics and quantum mechanics. Tools such as molecular dynamics simulations [8,9], quantum chemistry, and machine learning [10] have significantly expanded our ability to model complex biological systems. Despite these advances, many challenges remain. Molecules in living cells do not behave in isolation. They exist in crowded environments, often surrounded by water and other biomolecules, with strong polar interactions constantly influencing their behavior. This complexity makes accurate simulation very difficult, especially because these interactions are highly sensitive to both atomic positions and electronic states. To obtain reliable results, simulation models must operate at the atomic or even electronic level. However, that level of detail requires substantial computational power. Today's CPUs and GPUs, though powerful, are still limited when it comes to simulating large biological systems with high precision. Even when the computations are feasible, classical models often need extensive manual tuning to align with experimental data. This raises further concerns about their general reliability and predictive power.

One concrete example is signal transduction, a critical process in cellular biology that governs how cells respond to external stimuli through cascades of protein–protein and protein–ligand interactions [11]. Many signaling pathways begin when a protein binds to another protein or a ligand. This initial binding can trigger conformational changes, or in some cases, enzymatic reactions such as phosphate group transfers. After the interaction, the molecules separate and continue downstream signaling. Ideally, we would like to simulate this entire sequence, from the unbound to the bound state, within a realistic aqueous environment. In practice, though, current molecular dynamics methods typically require the protein and ligand to be positioned in nearly optimal binding orientations before the simulation begins. Allowing a ligand to drift freely in water and naturally bind to a protein during the course of a simulation is almost impossible with classical methods. The reason lies in computational limitations: simulation boxes cannot be too large, and timescales must remain short to be computationally manageable using CPUs or GPUs.

The challenge becomes even greater when electron transfer processes are involved, as in enzymatic catalysis or charge redistribution. These scenarios demand quantum chemical approaches. Hybrid QM/MM methods have been developed to combine quantum-level accuracy with the scalability of classical models [12]. Nevertheless, such methods remain computationally intensive and restricted in terms of system size and simulation time. The result is a trade-off between accuracy and feasibility. Ultimately, if we aim to simulate molecular systems with high precision and at biologically relevant scales, classical computing alone may not be sufficient. Quantum computing offers a fundamentally different paradigm. By working directly at the quantum level and enabling large-scale parallelism, it holds the potential to simulate complex molecular systems more naturally and efficiently. It could go beyond merely improving simulation fidelity and might eventually lead to new insights that challenge existing physical models based on classical approximations.

Another illustrative example is structure-based drug screening, where the goal is to identify ligands with strong binding affinity to a known protein target [13]. Given the 3D structure of the protein and a large ligand library, the task involves scanning for potential molecules that can stably and selectively bind to the protein's active site. To achieve accurate predictions, several critical factors must be considered: (1) the conformational flexibility of the protein, (2) the sheer scale of the ligand chemical space, which is estimated to contain over $10^{60}$ drug-like compounds , (3) the conformational variability of each ligand, (4) the spatial position and orientation of the ligand within the binding pocket, including translational and rotational degrees of freedom, and (5) the quality and accuracy of the scoring function used to evaluate binding affinity [14, 15]. Due to computational limitations on classical hardware, especially in large-scale simulations, many of these aspects are either simplified or ignored. Protein and ligand flexibility is often not fully accounted for, or only limited conformational sampling is performed. The number of ligands that can be realistically screened depends heavily on the available GPU resources. Regarding docking accuracy, a ligand placement within approximately 2 Å of the correct binding pose is typically accepted, although higher precision would require significantly more computation [16, 17]. Similarly, the scoring functions used are compromises between accuracy and efficiency, often relying on empirical approximations rather than quantum-chemical precision [18, 19]. The outcome of this trade-off is that the screening process rarely identifies the best possible ligands in terms of binding strength. However, it still offers practical value. By narrowing down the library to a subset enriched with likely binders, the chance of discovering strong hits during experimental validation is improved compared to random selection from the original library. In this sense, classical screening helps concentrate the chemical space, even if it does not pinpoint the optimal candidates. The core bottleneck lies in the vast combinatorial space involved in ligand selection, pose generation, and conformational sampling. These tasks are inherently parallelizable, making them well suited for QPUs. With the ability to explore multiple binding configurations simultaneously and evaluate complex quantum interactions directly, QPUs have the potential to address the computational challenges that currently limit the effectiveness of structure-based drug discovery on classical platforms.

In classical neural networks, learning is achieved by adjusting numerical parameters, such as weights and biases. In contrast, quantum neural networks rely on parameterized quantum gates, where one or more real-valued parameters control each gate. These parameters determine how quantum states evolve, and through training, the system learns to perform specific tasks based on the data. Over the past few years, numerous designs for quantum neural networks have been proposed [20-32]. However, one persistent challenge across many of these architectures is the so-called barren plateau problem [33]. The barren plateau refers to a situation where the gradient of the loss function becomes vanishingly small as the number of qubits or layers increases. When this happens, training the model becomes tough because the optimizer can no longer find a meaningful direction in parameter space. This issue tends to worsen in deeper circuits, especially those with high entanglement. To address this, researchers have explored various strategies. Some circuit designs aim to reduce unnecessary entanglement. Others rely on careful parameter initialization schemes that help avoid flat regions in the loss landscape [34].

Another critical challenge in quantum neural networks is the lack of intrinsic nonlinearity. In classical neural networks, nonlinear components such as activation functions and pooling layers play a central role in enabling the model to extract meaningful features and learn complex patterns from data. These nonlinear operations are essential for building deep and expressive architectures. However, quantum operations are inherently linear, which makes it difficult to replicate the nonlinear transformations that classical networks rely on directly. While quantum measurement introduces nonlinearity by collapsing a quantum state into classical outcomes, it lacks the control and expressive power of classical nonlinear functions. To simulate nonlinear behavior, some methods involve performing repeated measurements. However, these techniques typically require a large number of circuit executions, increasing the computational cost. Moreover, the measurement results must be processed on a classical computer and then, if further quantum processing is needed, re-encoded back into quantum states. This additional overhead reduces the efficiency and scalability of the model. One strategy to address this limitation is to approximate nonlinear functions using series expansions that are compatible with the linear nature of quantum operations [35, 36]. By expressing a nonlinear function as a sum of polynomial terms, it becomes possible to encode nonlinearity indirectly into the quantum model. Although this method expands the input space significantly, tensor products can be used to extract the relevant higher-order features efficiently. This approach offers a practical path for introducing nonlinear effects into quantum neural networks, without violating the mathematical constraints of unitary evolution.

In quantum circuits that use gates such as Rx, Ry, or Rz, even a small change to a single parameter can influence the entire output quantum state. This global sensitivity allows quantum neural networks to operate with fewer parameters than classical networks while still achieving accurate data fitting. However, several significant challenges remain. When too many parameterized gates are used, training can become unstable or stall entirely due to optimization difficulties. Additionally, nonlinear quantum gates do not exist within the standard framework of quantum computing, making it difficult to reproduce the expressive power of classical architectures. Designing circuits that are flexible enough to perform a wide range of mathematical tasks also remains a complex problem. These issues limit the performance and scalability of fully quantum models, which have led many researchers to explore hybrid approaches that combine classical and quantum components [37-39]. In light of these limitations and variations in performance, we propose a new quantum neural network architecture designed with five key objectives. First, it aims to mitigate the vanishing gradient problem that often arises during training. Second, it incorporates essential nonlinear effects to enhance learning capacity. Third, it supports a large number of parameters to allow for more flexible data fitting.

**Methods**

**Incorporating nonlinear effects using orthonormal bases.** In classical deep learning, each layer of a neural network processes the inputs by taking weighted sums and applying nonlinear activation functions. However, in quantum computing, we can't directly implement nonlinear functions with unitary matrices. To work around this, we apply Taylor series expansions to approximate these nonlinear behaviors. While the function itself is nonlinear in terms of the raw data, when it's rewritten in terms of the power series, it becomes linear in the basis terms. So if we first convert the raw data into these expanded basis terms and feed them into a quantum circuit, the resulting quantum states reflect nonlinear behavior about the original data.

The usual neural network operation can be described as:

$$f(X) = \sigma^s\left(W^s\left(\ldots\sigma^1\left(W^1 X + b^1\right)\ldots\right) + b^s\right) \quad (1)$$

Where $f$ is a nonlinear activation function applied to the weighted sum of inputs, but in a quantum circuit, we can't directly implement this because the operations must remain unitary. To address this, we express it as a Taylor series [40]:

$$f(x_0, x_1, \ldots, x_{n-1}) = a + a_{m_0}\sum_{m_0=0}^{n-1} x_{m_0}$$
$$+ a_{m_0 m_1}\sum_{m_0=0}^{n-1}\sum_{m_1=m_0}^{n-1} x_{m_0} x_{m_1} + a_{m_0 m_1 m_2}\sum_{m_0=0}^{n-1}\sum_{m_1=m_0}^{n-1}\sum_{m_2=m_1}^{n-1} x_{m_0} x_{m_1} x_{m_2} + \ldots \quad (2)$$

This expansion allows us to interpret nonlinear functions as linear combinations over a polynomial basis. By preparing quantum states that correspond to these basis functions, we can indirectly capture nonlinear behavior using linear quantum operations.

**Designing the quantum convolution architecture.** To prepare the quantum input states, we take neighboring data points and compute their pairwise products. This is shown in Figure 1a and 1b. Next, let's consider two example convolution kernels (Figure 1c). We first flatten each of these kernels into row vectors and then stack them together into a matrix (Figure 1d). As shown in Figures 1d and 1e, we compute the inner product between the convolutional kernels and the nonlinearly expanded input by multiplying the kernel matrix (Figure 1d) with the transformed data matrix (Figure 1e), which originates from the pairwise products in Figure 1b.

Quantum circuits need unitary matrices. So to ensure our matrix satisfies this condition, we use singular value decomposition (SVD). If the resulting U and V* matrices from SVD are the same size, we just multiply them together [41]. If they're different sizes, we pad the smaller one with 1s on the diagonal before multiplying.

We then build our quantum circuit by combining the convolutional kernel matrix with identity gates, as shown in Figure 1f [42, 43]. The output and input quantum states in this circuit are related through matrix operations, illustrated in Figure 1g and translated mathematically in Figure 1h. By turning the kernel into a unitary matrix, the inner product operation matches standard matrix multiplication, just like in Figures 1d and 1e.

We implemented these matrix operations and parameter updates using PyTorch [44]. Since both the MNIST and Fashion MNIST datasets require predicting one of 10 classes, we added a classical fully connected layer at the end to produce a one-hot encoded output [45, 46].

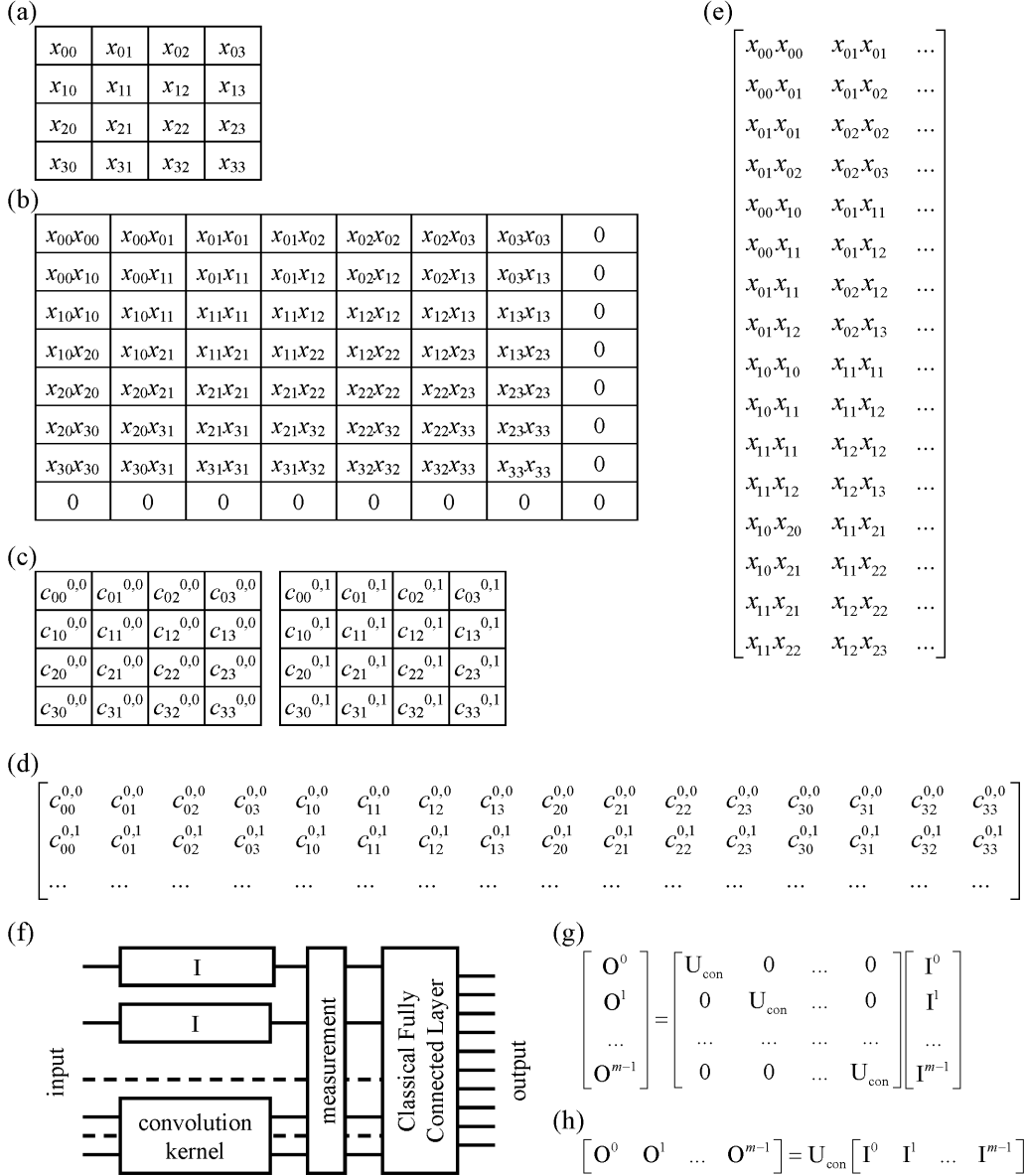

**Figure 1. Quantum Neural Network Architecture.** This figure illustrates the design

of a quantum convolutional neural network. (a) Shows a 4×4 input image represented by pixel values $x_{ij}$; (b) displays an expansion of the image where each entry is the product of adjacent pixels, introducing nonlinear features; (c) depicts multiple convolutional kernels; (d) shows all kernel vectors concatenated into a single matrix; (e) each column in the transformed data corresponds to a set of pixels from (b); (f) outlines the quantum circuit, consisting of convolution kernels, identity gates, quantum measurements, and a classical fully connected layer; (g) describes the output–input relationship using block-diagonal unitary matrices; and (h) reformulates this relationship as a matrix equation involving the unitary transformation of the kernels.

**Hyperparameters used for experiments.** We resized the original 28×28-pixel images in the datasets to both 8×8 and 32×32 using the Python Imaging Library. The 8×8 version is faster to compute, which helps especially when using quantum simulators like Qiskit. To incorporate nonlinear interactions, we expanded the input data using multiplication factors of 1×1, 2×2, 4×4, and 8×8. These factors determine how much of the surrounding pixel region is considered when computing pairwise products between neighboring pixels. For example, a 2×2 multiplication factor means that each transformed feature encodes the product interactions within a 2×2 local window of the original image, as illustrated in Figure 1b.

The size of the convolutional kernel is defined relative to this expanded representation. In this study, we used nominal kernel sizes of 1×1, 2×2, 4×4, and 8×8, referring to the number of spatial blocks spanned within the multiplication-expanded feature map. However, the actual matrix dimensions of the convolutional kernel are determined by multiplying the kernel size with the multiplication factor along each spatial axis. For instance, when a 2×2 kernel is applied to data expanded by a 1×1 multiplication factor, the resulting kernel matrix has an effective size of 2×2. When the multiplication factor increases to 2×2, 4×4, or 8×8, the corresponding actual kernel dimensions become 4×4, 8×8, and 16×16, respectively. We fixed the stride of the convolutional kernels to 2 and set the number of kernels to 16 in all configurations.

Another important hyperparameter in our training setup is the learning rate used in stochastic gradient descent (SGD) with momentum. To determine a suitable value, we tested four learning rates: $10^{-4}$, $10^{-3}$, $10^{-2}$, and $10^{-1}$. Each learning rate was evaluated by running the model three times using independently randomized parameter initializations. During each run, parameters were initialized from a uniform distribution and trained using SGD with a fixed momentum coefficient of 0.9. For robustness, we recorded the lowest training accuracy across the three runs for each learning rate. The learning rate corresponding to the highest of these minima was selected, and the associated model parameters were used to report the final training and test accuracy.

**Verification of Quantum Simulation Consistency.** To verify whether the PyTorch-based implementation effectively simulates the intended quantum operations, we used the model parameters corresponding to the lowest training accuracy from the learning rate selection experiments. These parameters were then applied to construct equivalent quantum circuits using Qiskit. By replicating the same circuit structure, input encoding, and unitary transformations within the Qiskit framework, we simulated the exact inference process and compared the predicted outputs. This step serves to validate the correctness of the PyTorch simulation by ensuring that the resulting probabilities and classifications are consistent with those obtained from a genuine quantum circuit simulation.

**Results**

**Quantum Convolutional Model Accuracy Assessment.** To generate nonlinear effects in the quantum convolutional model, we used the products of input pixels. Since it's computationally expensive to include all possible pairwise or higher-order combinations, we chose to represent these effects using specific multiplication factors: (1×1), (2×2), (4×4), and (8×8). Each factor determines how many surrounding pixels are considered when computing the product terms. Larger multiplication factors introduce richer nonlinear interactions between pixels. Another critical parameter is the convolutional kernel size, which is known to affect the accuracy of convolutional neural networks significantly. In this study, we used kernel sizes of (1×1), (2×2), (4×4), and (8×8). To avoid excessive combinations of hyperparameters, we fixed the kernel stride to 2. While increasing the number of convolutional kernels can improve model accuracy, we kept the number of kernels fixed at 16 across all experiments for consistency. The model architecture we used corresponds to the quantum convolutional network shown in Figure 1f. We tested this setup using the MNIST and Fashion-MNIST datasets. In both cases, we resized the original 28×28-pixel images to either 32×32 or 8×8 to match different experimental conditions.

For the MNIST dataset, the same quantum model yielded consistent accuracy across both training and test sets. The best result was observed when the multiplication factor and the kernel size were both set to (8×8), giving us a test accuracy of 99.0%, as shown in Table 1. On the other hand, the results for Fashion-MNIST showed a more noticeable difference between the training and testing accuracies. In that case, the highest test accuracy was 88.0%, also achieved using the (8×8) multiplication factor and the (8×8) kernel size. Detailed numbers for both datasets are provided in Table 1 and Table 2, respectively.

**Table 1. MNIST dataset accuracy.** This table summarizes the training and test accuracies of the quantum convolutional network across different image sizes, convolutional kernel sizes, and input data multiplication factors. For each combination, the model's accuracy is reported separately for training and testing.

| image sizes | size of the convolution kernel | multiplication factor | | | | | | | |
|---|---|---|---|---|---|---|---|---|---|
| | | 8 * 8 | | 4 * 4 | | 2 * 2 | | 1 * 1 | |
| | | train | test | train | test | train | test | train | test |
| 32 * 32 | 8 * 8 | 99.6 | 99.0 | 99.3 | 98.8 | 99.1 | 98.5 | 97.8 | 97.0 |
| | 4 * 4 | 99.4 | 98.9 | 98.8 | 98.3 | 98.4 | 97.8 | 97.0 | 96.2 |
| | 2 * 2 | 98.9 | 98.3 | 98.4 | 97.6 | 96.7 | 96.0 | X | X |
| | 1 * 1 | 98.6 | 98.0 | 97.1 | 96.1 | X | X | X | X |
| 8 * 8 | 8 * 8 | 98.9 | 98.1 | 98.4 | 98.0 | 97.1 | 97.0 | 92.6 | 92.8 |
| | 4 * 4 | 98.9 | 98.2 | 98.0 | 97.8 | 96.4 | 96.5 | 92.4 | 92.7 |
| | 2 * 2 | 97.9 | 97.6 | 96.7 | 96.7 | 93.5 | 93.3 | X | X |
| | 1 * 1 | 95.9 | 95.9 | 92.7 | 93.2 | X | X | X | X |

**Table 2. Fashion-MNIST dataset accuracy.** Similar to Table 1, this table shows the training and test accuracies of the quantum convolutional network, but applied to the Fashion-MNIST dataset.

| image sizes | size of the convolution kernel | multiplication factor | | | | | | | |
|---|---|---|---|---|---|---|---|---|---|
| | | 8 * 8 | | 4 * 4 | | 2 * 2 | | 1 * 1 | |
| | | train | test | train | test | train | test | train | test |
| 32 * 32 | 8 * 8 | 91.6 | 88.0 | 91.0 | 87.4 | 90.1 | 86.5 | 88.7 | 84.9 |
| | 4 * 4 | 91.5 | 87.3 | 90.5 | 86.7 | 89.2 | 85.5 | 86.8 | 83.1 |
| | 2 * 2 | 90.9 | 86.8 | 89.3 | 85.7 | 86.1 | 82.2 | X | X |
| | 1 * 1 | 89.9 | 85.8 | 86.8 | 83.1 | X | X | X | X |
| 8 * 8 | 8 * 8 | 89.8 | 87.1 | 88.7 | 86.6 | 87.1 | 85.4 | 82.9 | 81.6 |
| | 4 * 4 | 89.5 | 87.2 | 86.7 | 85.3 | 84.6 | 83.0 | 80.9 | 79.3 |
| | 2 * 2 | 87.9 | 85.8 | 84.9 | 83.5 | 81.7 | 80.1 | X | X |
| | 1 * 1 | 85.8 | 84.4 | 81.8 | 80.6 | X | X | X | X |

**Impact of Input Data Multiplication Factor on Accuracy.** Figure 2 illustrates how the input data multiplication factor affects classification accuracy across various datasets and convolutional kernel sizes. The figure consists of four subplots: (a) MNIST resized to 32×32, (b) MNIST resized to 8×8, (c) Fashion-MNIST resized to 32×32, and (d) Fashion-MNIST resized to 8×8. In each subplot, the x-axis indicates different convolutional kernel sizes, denoted as 'C', while the colored bars represent different multiplication factors: blue for (1×1), orange for (2×2), gray for (4×4), and yellow for (8×8).

Overall, increasing the multiplication factor tends to enhance model performance. The most significant improvement typically occurs when increasing the factor from (1×1) to (2×2). Further increases to (4×4) continue to improve accuracy, albeit with diminishing returns. In many cases, expanding to (8×8) yields only marginal gains, particularly when the accuracy under the (4×4) setting is already near optimal (e.g., close to 99%). However, when the accuracy under (4×4) remains relatively low, using (8×8) can still lead to notable improvements.

These results suggest that incorporating additional nonlinear interactions via pixel-wise multiplication enables the model to learn more complex and expressive patterns. This strategy is particularly effective when the initial configuration has not yet saturated the model's capacity, allowing further expansion to yield measurable accuracy gains.

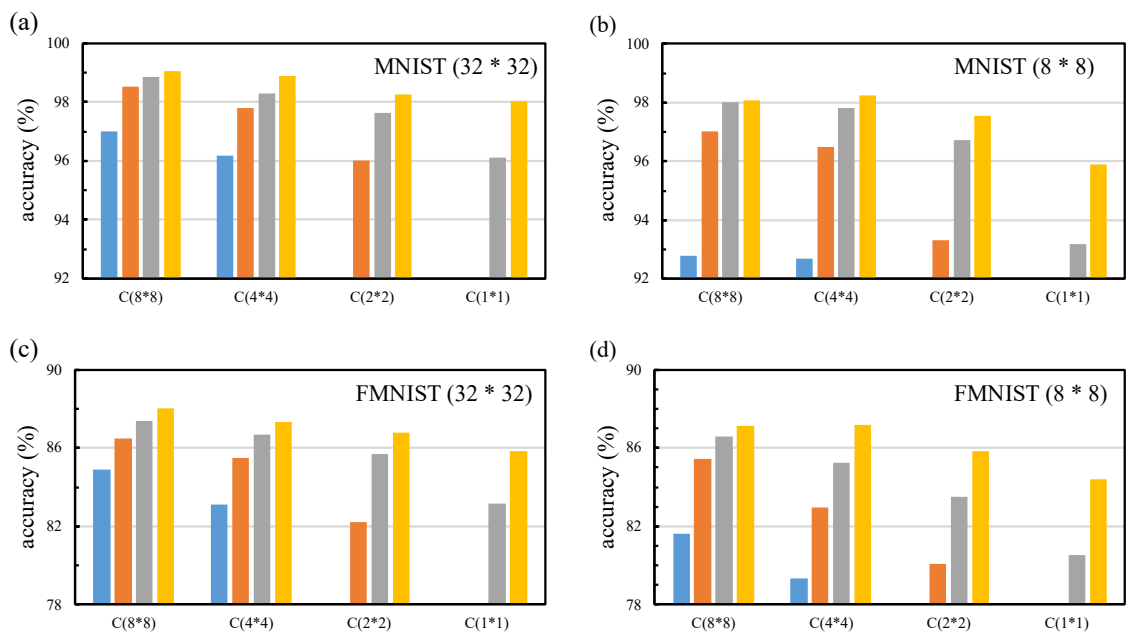

**Figure 2. Impact of Input Data Multiplication Factors on Accuracy.** The figure illustrates how different input data multiplication factors affect classification accuracy across various convolution kernel sizes. Each subplot corresponds to a different dataset setting: (a) MNIST resized to 32×32, (b) MNIST resized to 8×8, (c) Fashion-MNIST resized to 32×32, and (d) Fashion-MNIST resized to 8×8. The x-axis denotes convolution kernel sizes (e.g., C(8×8)), while the y-axis shows classification accuracy. Each color-coded bar represents a specific input multiplication factor: blue for 1×1, orange for 2×2, gray for 4×4, and yellow for 8×8.

**Impact of Convolutional Kernel Size on Accuracy.** In classical convolutional neural networks, kernel size significantly influences model performance. A similar trend is observed in our quantum convolutional model. Figure 3 presents results across different input multiplication factors and datasets, indicating that increasing the kernel size generally improves classification accuracy, particularly when starting from smaller kernels such as (1×1) or (2×2).

In most configurations, switching to a (4×4) kernel produces a noticeable improvement. The (4×4) kernel provides consistently strong and stable results across both the MNIST and Fashion-MNIST datasets, regardless of whether the input images are resized to 32×32 or 8×8. In some cases, further increasing the kernel size to (8×8) leads to slightly higher accuracy, although the gain is often limited. There are also instances where the performance of the (8×8) kernel is comparable to that of the (4×4), suggesting that the benefit of increasing kernel size may eventually saturate.

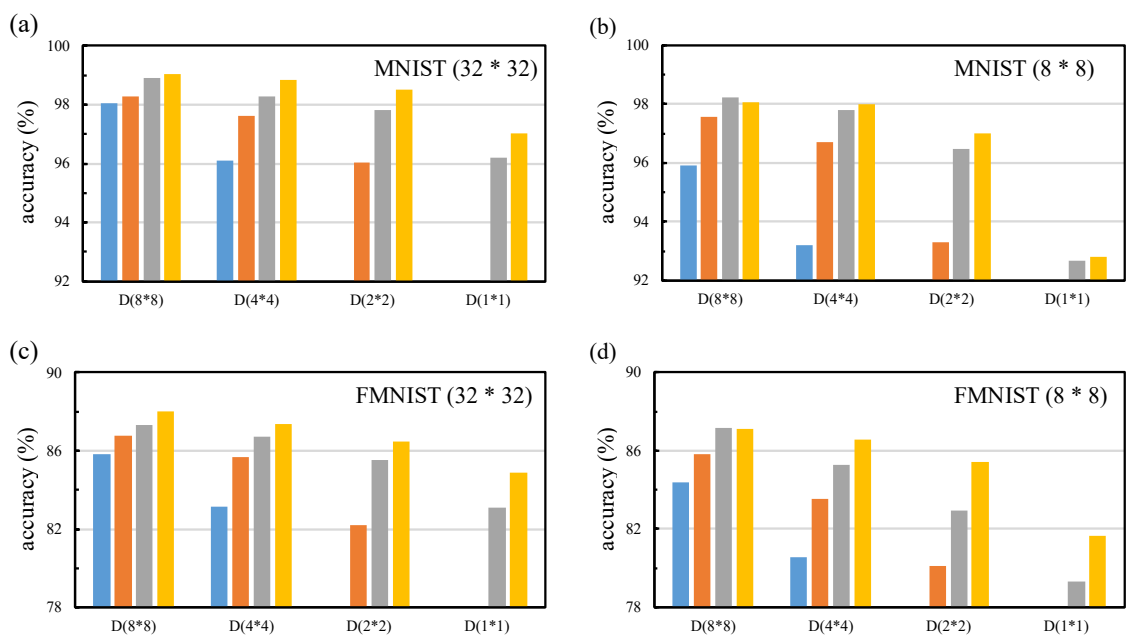

**Figure 3. Impact of Convolutional Kernel Size on Accuracy.** This figure shows how varying convolutional kernel sizes affect classification accuracy under different input data multiplication factors. Each color-coded bar corresponds to a kernel size: blue for 1×1, orange for 2×2, gray for 4×4, and yellow for 8×8. The x-axis indicates the input data multiplication factor (e.g., D(8×8)), while the y-axis shows test accuracy. Subplots represent results on four dataset settings: (a) MNIST resized to 32×32, (b) MNIST resized to 8×8, (c) Fashion-MNIST resized to 32×32, and (d) Fashion-MNIST resized to 8×8.

**Verification Using a Quantum Simulator.** To make training computationally feasible, especially under hardware limitations, we adopted the matrix-based approach illustrated in Figure 1h to implement and optimize our model using PyTorch. This method allowed us to simulate the quantum circuit efficiently using matrix operations, which significantly reduced memory usage and computation time during parameter training.

To ensure that the trained parameters would remain valid when applied to an actual quantum setting, we conducted a verification step using the Qiskit quantum simulator. In this process, we initialized quantum states through Qiskit's classical data encoding interface, applied the same trained unitary matrices using Qiskit's Unitary function, and then used Qiskit's internal probability calculation to extract the output distribution.

The comparison between PyTorch and Qiskit results is shown in Figure 4, where we evaluated the classification accuracy using both methods on the MNIST dataset resized to 8×8 pixels. The results from PyTorch and Qiskit are nearly identical, demonstrating that the learned parameters can be faithfully translated into a real quantum circuit and still yield consistent performance. This confirms that the PyTorch-based training procedure offers not only computational efficiency during simulation but also accurate and transferable results when deployed within an actual quantum computation framework.

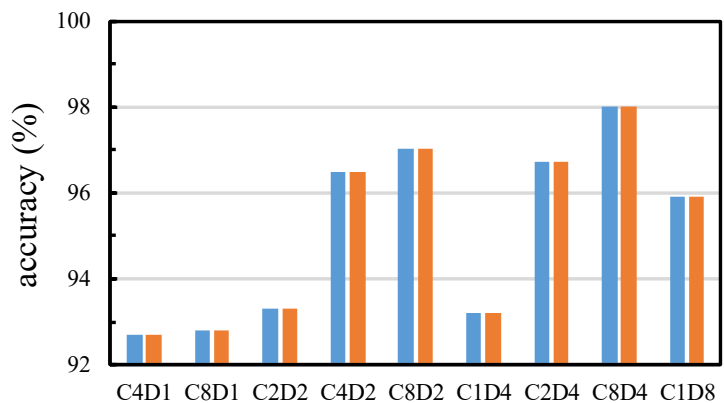

**Figure 4. Comparison of Accuracies Using PyTorch and Qiskit (MNIST resized to 8×8).** This figure compares test accuracies obtained using PyTorch (blue) and Qiskit's probability-based approach (orange). The x-axis labels indicate configurations where 'C' refers to the convolutional kernel size and 'D' to the input data multiplication factor (e.g., C8D4 means kernel size 8×8 with multiplication factor 4×4). The y-axis shows the classification accuracy.

## Discussion

**Introducing Nonlinear Effects.** In quantum circuits, the output state is always a linear combination of the input states. The only form of nonlinearity naturally available comes from the act of measurement, which converts amplitudes into probabilities. However, this is not sufficient to replace the role of nonlinear functions like activations in classical neural networks. To overcome this, we introduced an orthogonal basis framework that allows us to represent nonlinear functions as linear combinations over that basis. In this study, we used polynomial functions as the basis, since quantum tensor operations naturally produce quadratic and higher-order interactions. As shown in Figure 2, even including just a few quadratic terms already led to improved accuracy. Although it would be ideal to include all higher-order terms for a more complete representation, such an approach is computationally too demanding. Therefore, we limited our expansion to a manageable level for practical implementation.

**Avoiding the Barren Plateau in Quantum Training.** Quantum machine learning models rely on parameterized gates to learn from data. These gates often use trigonometric functions such as cosine, and their compositions through tensor and matrix products generate highly complex, non-convex landscapes. As the number of gates increases, the output state becomes a sum of many such cosine-based products. This can easily result in a loss function that is filled with local minima and nearly flat regions, making gradient-based training extremely difficult. To address this, we chose to parameterize entire unitary matrices instead of stacking multiple parameterized gates. After training the parameters in matrix form, we then converted the result into the corresponding quantum circuits.

**Quantum Convolutional Neural Network Architecture.** The architecture we propose is visualized in Figure 1f and is built using parameterized unitary matrices that act as quantum equivalents of convolutional kernels. Once training is complete, the values of these matrices can be substituted directly into a quantum circuit.

**Validating PyTorch Results with a Quantum Simulator.** To ensure that the PyTorch-based matrix implementation faithfully represents a real quantum system, we validated the results using a quantum simulator. We used the circuit shown in Figure 1f for training, but this circuit is too large to simulate repeatedly with high efficiency. Therefore, we trained the model using the simplified formulation in Figure 1h, and then checked that the resulting parameters still behave correctly when inserted into a full quantum simulation. The quantum simulator was used to verify fundamental conditions such as unitarity of the transformation matrix and the normalization of the output state. After training, we compared the prediction accuracy from both PyTorch and Qiskit. The agreement between them confirms that our method is valid and that the simplified training approach does not compromise physical consistency.

**Influence of Multiplication Factors and Kernel Sizes.** Due to hardware constraints, it was not feasible to include all possible pairwise products in the input representation. Instead, we systematically investigated the effects of different multiplication factors. The results demonstrate that increasing the multiplication factor significantly enhances model performance by introducing additional nonlinear interactions. This confirms that nonlinear input expansion plays a central role in boosting the expressiveness of the quantum network.

Regarding kernel size, the experimental results also show a general trend: larger convolutional kernels tend to improve accuracy. However, when comparing their relative impact, the multiplication factor exerts a much stronger influence on performance than kernel size. This suggests that enriching the input data through nonlinear transformations is more effective for model learning than simply enlarging the spatial coverage of the convolution.

**Conclusions**

In this study, we proposed a quantum convolutional neural network architecture that incorporates nonlinear effects and avoids the barren plateau problem, which often impedes the training of quantum machine learning models. By using orthonormal polynomial bases to approximate nonlinear functions and by directly parameterizing unitary matrices instead of stacking individual gates, we enabled the model to learn expressive features while maintaining computational tractability.

The architecture was tested on both the MNIST and Fashion-MNIST datasets, achieving high classification accuracy. Results also demonstrated that larger input multiplication factors, which introduce more nonlinear interactions, have a greater influence on performance than the convolutional kernel size alone. In addition, we verified the consistency of predictions between classical PyTorch simulations and quantum circuit simulations using Qiskit, which confirmed the physical validity of the trained parameters.

Overall, the proposed framework provides a flexible, scalable, and interpretable approach to quantum convolutional learning. It bridges the gap between classical convolutional architectures and quantum computing principles, offering a promising direction for future research in hybrid and fully quantum machine learning systems.

**Data and Software Availability**

The data supporting the findings of this study are openly available on GitHub at the following URL: https://github.com/peikunyang/03_activation.